\documentclass[doublecol]{epl2} 

\newcommand{\avg}[1]{\langle{#1}\rangle}

\def\be{\begin{equation}}
\def\ee{\end{equation}}
\def\bc{\begin{center}} 
\def\ec{\end{center}}
\def\bea{\begin{eqnarray}}
\def\eea{\end{eqnarray}}
\def\bwt{\begin{widetext}}            
\def\ewt{\end{widetext}}

\title{Non-neutral theory of biodiversity}
\shorttitle{Non-neutral theory of biodiversity} %Insert here a short version of the title if it exceeds 70 characters

\author{Ginestra Bianconi\inst{1}, Luca Ferretti\inst{2} and  Silvio Franz\inst{3}}
\shortauthor{G. Bianconi \etal}

\institute{                    
  \inst{1} The Abdus Salam International Center for Theoretical
  Physics, Strada Costiera 11,34014 Trieste,Italy\\
  \inst{2}  Facultat de Veterinaria and IFAE,
  Universitat Autonoma de Barcelona, 08193 Bellaterra, Spain
\inst{3}  LPTMS, UMR 8626 CNRS et Universit\'e Paris-Sud, B\^atiment
100, 91405 Orsay CEDEX, France
}

\abstract{
We present  a  non-neutral stochastic model for the dynamics taking
place in a meta-community ecosystems in presence of migration. The model provides a framework for describing the
  emergence of  multiple   ecological scenarios and behaves  in  two
  extreme limits either  as the unified neutral theory of biodiversity or as the Bak-Sneppen
  model. Interestingly, the model shows a condensation phase transition where one
species becomes the dominant one, the diversity in the ecosystems is
strongly reduced  and the  ecosystem is 
non-stationary. This phase transition extend the principle of
competitive exclusion to open ecosystems and might be relevant for the
study of the impact of invasive species in native ecologies.
 }

\begin{document}

\maketitle

Understanding  the basic principles responsible for the 
biodiversity in ecosystems is a fundamental questions of scientific
and social interest. In fact many   ecosystems recently, undergo   
 a loss of biodiversity \cite{marine,invasive} due to
human activities or to non-native invasive species that in some cases
might become dominant in the ecosystem.

Recently the {\it unified neutral theory of biodiversity and biogeography}
\cite{Hubbell} has been shown
to reproduce species abundance distributions   \cite{McKane,Maritan0} of meta-communities and
local communities and the  distribution of
lifetimes  \cite{Pigolotti} that are found in ecological systems.
The proposed theoretical framework is based on the assumption that all
species are equivalent and  assumes  that species abundance
distribution is  affected mainly  by the drift resulting from a
stochastic process of  births, deaths and speciation.
This theory raised some controversies  in the field \cite{McGill,reef}, and there is an increasing need of a theory
which includes the stochastic features of the neutral model but also
accounts for the competitive advantage that some species might have in
their niche.
Moreover a  non-neutral theory of meta-communities biodiversity has the
potential to include also complex prey-predator foodweb relations
\cite{Cohen,Martinez,Dunne} for which  an increasing number of data is
now available.

In the physics community non-neutral ecological models have been widely studied
 \cite{Dunne,Drossel,Drossel1}. In this context a reference model is   the Bak-Sneppen model \cite{BS,BSF}, which  
describes the extinction process in a non-neutral ecosystem. In the
model, a negative selection process takes place and at each step  the less fit species of the ecosystem is removed,
and  mutations affect the  species dependent on the removed one. This model  leads to
punctuated equilibrium and power-law extinction patterns found in
fossil records.

In this paper we show that these two paradigms (the neutral model and the
Bak-Sneppen model) can be recovered by a
simple model of metacommunities ecological  systems  \cite{Dunne,Drossel}  evolving following the
principles of evolutionary dynamics, which include
the fundamental ingredients  of reproduction, negative selection and migration/speciation.
In our stochastic  model we assume that
species are not equivalent but they are characterized by a quenched
parameter that encodes for the adaptation of the species to their niche.
We call this parameter the fitness of the species. High fitness will
imply a competitive advantage of the species as in evolutionary models. In particular high fitness will increases
the probability of reproduction of a species, and reduce the death
probability due to a  negative selection process. In the model
speciations processes are included and we assume that they occur mainly
as a consequence of migration events, in order to describe  invasive species arriving at fixed rate in an ecosystem.  
This model reduces to the neutral theory of biodiversity in the limit in which all species have the same fitness. 
On the other side when we include the effect of evolutionary selection
on species of
heterogeneous fitness  the model has a rich
phenomenology. In the limit in which  only negative selection is
present, and considering an extremal dynamics, we recover a dynamics
that can be mapped into the 
mean-field Bak-Sneppen model  \cite{BS,BSF}.
Moreover the model shows a  
phase transition to a non-stationary state
in which one species becomes the dominant species in the ecosystem and
the biodiversity of the ecosystem is strongly reduced. 
This phase transition is a condensation phase transition that  in
some limiting cases can be mapped to a Bose-Einstein condensation and is closely related to the
condensation transitions found in different evolving models
 \cite{Kingman,Franz,Bose,Maritan1,other2}.
Below this phase transition, relevant dynamical effects take place and
the condensed species  with a finite population change in time in a
nontrivial way.

{\it  Non-neutral birth-death dynamics of the ecosystem -}
We generalize the stochastic birth-death dynamics giving
rise to the distribution of species in meta-communities in the neutral
theory of biodiversity to a birth-death dynamics depending on quenched
variables assigned to the species and that correspond to their niche
adaptation. We call this quenched variable the fitness $f$ of a species.
We consider an ecosystem formed by $N$ individuals belonging to  $S$ species with the
number of individuals $N$ fixed but a variable number of species $S$. 
Each individual  belongs to a species
$i=1,2,\ldots,S$ and   each species $i$ has fitness $f_i>0$.
The birth-death dynamics is inspired by evolutionary dynamics and
three different birth-death processes take place. We call this
three processes replication, negative selection and migration/speciation. 
Starting from a random initial condition we define the following
birth-death process:
\begin{itemize}
 \item{\bf Replication -}
At  rate $\alpha$ an  individual of the species $i$ with fitness $f_i$
is chosen to replicate with probability 
 $w(f_i)$, where $w(f)$ is an increasing function of the fitness,   and is substituted by a random individual of the
 ecosystem.
Therefore the resulting process is 
\bea
n_i\rightarrow n_i+1&\mbox{and }& n_j\rightarrow n_j-1
\eea
and the species $i$ and
$j$ are selected according to the probabilities $\Pi_i^R$ and $\Pi_j^R$ respectively, defined as
\bea
\Pi_i^R=\frac{w(f_i) n_i}{\sum_{\ell} w(f_{\ell})n_{\ell}}&\mbox{and }
& \Pi_j^R=\frac{n_j}{N}.
\eea
This is the usual process of reproduction in population dynamics in
which $w(f_i)$ describes the mean number of expected off-springs of
the individual $i$.

\item{\bf Negative selection -}
At  rate  $\gamma$ an individual of a species $i$ with
fitness $f_i$ is removed from the population with probability
$B(f_i)$, which is a decreasing function of the fitness $f_i$ and is substituted by an offspring of a random
individual.
The resulting process is 
\bea
n_i\rightarrow n_i-1&\mbox{and }& n_j\rightarrow n_j+1
\eea
and the species $i$ and
$j$ are selected according to the probabilities $\Pi_i^{NS}$ and
$\Pi_j^{NS}$ respectively, defined as
\bea
\Pi_i^{NS}=\frac{n_i B(f_i)}{\sum_{\ell} B(f_{\ell})n_{\ell}}
&\mbox{and }&\ \Pi_j^{NS}=\frac{n_j}{N}.
\eea
This is a term of negative selection  and  describes the struggle for
survival of the individuals. In particular $B(f_i)$ we take
\be 
B(f_i)=1/w(f_i)
\ee  describing  the Darwin concept of removal of
the less fit species. More general form of $B(f)$ might instead
represent barrier between fitness peaks in some  complex fitness
landscape.
 
This type of process is the pivotal process of the Bak-Sneppen  model
which is in addition also extremal (at each time-step the less fit
species is removed from the ecosystem {or migrates away from
  the metacommunity }).
We observe here that   in our model when $\alpha=0$, $\gamma<1$ and
$\beta\rightarrow \infty$, we recover an extremal dynamics that is
reminiscent of the Bak-Sneppen model\cite{BS,BSF}.

One might naively think that the effect of the negative selection
process and of the reproduction process is equivalent. This is not the
case in general as it will be shown in the following.
 
\item{\bf Migration/Speciation} 
At rate  $\mu=1$ a migration/speciation occurs and a random 
individual is substituted  with an individual of a new species  $i=S+1$. The new
species is assigned a   fitness $f_{S+1}$ randomly drawn from a
fixed  distribution $\rho(f)$. The resulting process is 
\bea
n_{S+1}\rightarrow 1&\mbox{and }&n_j\rightarrow n_j-1
\eea
where the species $j$ is selected with probability 
$\Pi^M_j=\frac{n_j}{N}$ and the fitness $f_{S+1} $ is drawn from a
distribution $\rho(f)$.
We consider here on purpose only the case  in which the new species
have fitnesses independent on the fitnesses of the species present in
the ecosystem. This is the case  of speciations due to migration from
another metacommunity which is relevant for the study of invasive
species in given ecosystems. 
\end{itemize}

In particular we consider the case in which  the function 
\be
w(f_i)=e^{\beta f_i}.
\label{ass}
\ee
where $\beta$ is a tunable parameter.The assumption $(\ref{ass})$ is by no means a limitation since
the model could be defined equivalently only in terms of the distribution of the
variables $\{w\}$'s.
Moreover the  choice of the assumption $(\ref{ass})$ has  the
following advantages:
\begin{itemize}
\item
i) It  ensure that we have always $w(f_i)>0$.
\item
ii) It allows to mimic the impact of the environment in driving the system
out of neutrality, by changing $\beta$.
In fact   $\beta$ tunes the relevance of the fitness in the
stochastic dynamics. In particular when $\beta=0$ we have $w_i=1\
\forall i \ $ and in the dynamical process all species are equivalent, recovering  the neutral birth-death model.  
\end{itemize}

{\it Solution of the model -}
We study the master equation of the  stochastic process describing the
non-neutral dynamics taking place in the model ecosystem.  
If we call $N_n(f,t)$ the number of  species
of  fitness $f$  populated by $n$ individuals,
its master equation is given by  
\bea
\frac{d N_n(f,t)}{dt}=b_{n-1} N_{n-1}(f,t)-b_{n} N_{n}(f,t)+\nonumber \\
\hspace*{2mm}d_{n+1} N_{n+1}(f,t)-d_n N_n(f,t)+ N\rho(f)\delta_{n,1},
\label{master}
\eea
where the birth rates $b_n$ and       
death rates $d_n$ are given by 
\bea
b_n(f)&=&\frac{n}{N}B(f)-\left(\frac{n}{N}\right)^2 C(f) \nonumber \\
d_n(f)&=&\frac{n}{N}D(f)-\left(\frac{n}{N}\right)^2C(f)
\eea
with
$B(f)=\alpha {e^{\beta f}}/{Z_1}+\gamma$,
$D(f)=1+\alpha+\gamma {e^{-\beta f}}/{Z_2}$
and $C(f)=\alpha {e^{\beta f}}/{Z_1}+\gamma {e^{-\beta f}}/{Z_2}$ 
and where we have defined $Z_1,Z_2$ as
\bea
{Z}_1(t)&=&\frac{1}{N}\int df \sum_n e^{\beta f_i}nN_n(f,t)\\
\nonumber
{Z}_2(t)&=&\frac{1}{N}\int df \sum_n e^{-\beta f_i}nN_n(f,t).
\eea
In order to solve the master equation $(\ref{master})$
we assume that $Z_1(t)$ and $Z_2(t)$ in the asymptotic limit
$t\rightarrow \infty$
 converge to a constant value 
\bea
{Z_{1/2}}(t)\rightarrow {\cal Z}_{1/2}+{\cal O}(N^{-1/2})
\eea
In this assumption, the steady state of the master equation
$(\ref{master})$ can be solved exactly as in neutral birth-death models  \cite{McKane,Maritan0} giving the distribution of the abundance of the species with
given fitness. The stationary solution for  $N_n(f)$ is given by
\be
N_n(f)=N\frac{\rho(f)}{d_1(f)}\prod_{i=2}^n \frac{b_{i-1}(f)}{d_i(f)}
\ee
In the limit in which $n\ll N$ this distribution becomes a convolution
of different Fisher log series distribution  \cite{Maritan0}
\be
N_n(f)=N\frac{\rho(f)}{D(f)}\frac{[\theta(f)]^{n-1}}{n}
\label{LF}
\ee 
with the diversity  number $\theta=\theta(f)$ dependent on the fitness
of the species $f$ according to the relation 
\be
\theta(f)=B(f)/D(f)=\frac{\alpha {e^{\beta f}}/{\cal Z}_1+\gamma}{\gamma
  {e^{-\beta f}}/{\cal Z}_2+(\alpha+1)}.
\label{theta}
\ee

From now on we focus on this model in the case of bounded fitness
distributions with  $f\in[f_m,f_M]$ for which we might expect that, al
least in a certain region of the phase space we might have a
stationary state of this evolutionary dynamics. Without loss of
generality we take from here on  $f\in [0,1]$.
In this case  the diversity parameter
$\theta(f)$ reaches a maximum at the maximal fitness $f=1$.
The average number of individuals with a given fitness $f$ is 
\be
\avg{n(f)}=N\frac{\rho(f)}{1+\alpha-\gamma+
\gamma  {e^{-\beta f}}/{\cal Z}_2-\alpha {e^{\beta f}}/{\cal Z}_1}
\label{average}
\ee
This distribution admits two relevant limits
$\alpha=0$ and $\gamma=0$. To show this, it is useful to put $f=1-\varepsilon$ and to define a
distribution $g(\varepsilon)=\rho(1-\varepsilon)$.
In absence of negative selection, i.e. $\gamma=0$, the average
number of individuals of fitness $f=1-\varepsilon$, $\avg{n(f)}$  can be expressed
in terms of a Bose-Einstein distribution of the occupation of the energy
level $\varepsilon$.

Conversely when reproduction selection are absent, 
i.e. $\alpha=0$, the average number of species $\avg{n(f)}$ with
fitness $f=1-\varepsilon$   follows a Fermi-Dirac occupation of the
energy levels $\varepsilon$, provided that the
negative selection processes occur at a smaller rate than 
speciation, i.e. when $\gamma<1$. {This show clearly that the distribution
of species in this evolutionary  dynamics change significantly if
only the reproduction selection process occurs or if only to the negative
selection process occurs.} 
Moreover taking the limit $\beta \rightarrow \infty$ for  $\alpha=0$, $\gamma<1$,   we recover for $\avg{n(f)}$ the step function
characteristic of the Bak-Sneppen model \cite{BS} and we can map the
dynamics to the dynamics of a mean-field Bak-Sneppen model \cite{BSF} showing
punctuated equilibrium.

Eq. $(\ref{LF})$ and Eq. $(\ref{average})$ are well defined once ${\cal Z}_1$ and ${\cal Z}_2$
are found self-consistently.
The self-consistent equations read 
\bea
{\cal Z}_1=\int_0^1 d\varepsilon g(\varepsilon)\frac{e^{-\beta \varepsilon}}{1+\alpha-\gamma+
\gamma  {e^{\beta \varepsilon}}/{\cal Z}_2-\alpha {e^{-\beta \varepsilon}}/{\cal Z}_1}\nonumber \\
{\cal Z}_2=\int_0^1 d\varepsilon g(\varepsilon) \frac{e^{\beta \varepsilon}}{1+\alpha-\gamma+
\gamma  {e^{\beta \varepsilon}}/{\cal Z}_2-\alpha {e^{-\beta \varepsilon}}/{\cal Z}_1}.
\label{selfc}
\eea

These self-consistent equations have to be solved for each given
parameter value $(\alpha, \gamma,\beta)$ and each distribution of the
fitness $\rho(f)=g(\varepsilon=1-f)$. 
Eqs. $(\ref{selfc})$ might lack a solution when
the maximal fitness value is reached in the system at infinite time,
i.e.
when $\lim_{f\rightarrow f_M}\rho(f)=\lim_{\varepsilon\rightarrow 0}
g(\varepsilon)=0$.
In this case, a inverse critical temperature can be evaluated \cite{prossimo} such
that for $\beta<\beta_c$ the Eqs. $(\ref{selfc})$ have a solution while 
for $\beta>\beta_c$ they do not.
 This phase transition  can be mapped in the case
$\gamma=0$ to a Bose-Einstein condensation phase transition in a Bose gas.
As in the Bose-Einstein condensation the lowest energy state gets a
finite occupation, in this model  when $\beta<\beta_c$ we observe  the emergence of   dominant species in the ecosystem having a    finite
fraction of the total population. This dominant species can change with time
 and the ecological system is non-stationary.
{We observe here that the time needed for an increase in the average
fitness of the ecosystem in the condensed state  increase
with time in a multiplicative way. In fact we can say that this
evolutionary  model show in this limit relevant aging effects.
This  aging phenomena might be quite unrealistic for real
ecologies. This feature of the model can be overcome by including
genetic mutations that can occur on very long time scales. Genetic
mutations are therefore  negligible if we 
consider an ecosystem  which reach quickly the  steady states, but they  can be
have relevant effects on the time scales of the aging dynamics of the
ecosystem in the condensed phase. }

\begin{figure}
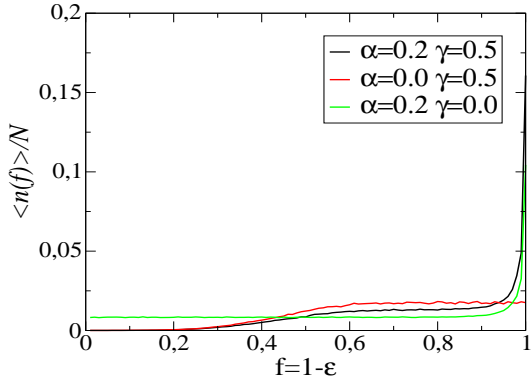

\onefigure[width=70mm,height=50mm]{dis.eps}
\caption{Fraction of the individuals of the population with fitness
  $f=1-\varepsilon$ for a uniform distribution $\rho(f)$ 
  with $f\in [0,1]$. The data are shown for  $\beta=20$
  and are averaged over   $100$ runs of the non-neutral birth-death
  dynamics involving a population of $N=1000$ individuals. The data
  are collected  at the time  $T=5\times 10^4$ of the stochastic dynamics. }
\label{dis.fig}
\end{figure}

{\it  Numerical characterization of the stationary and non-stationary
  state of the ecosystem -}
We have simulated the model and studied the
population distribution and collected numerical 
evidence for the condensation phase transition to a non-stationary state.
In fig. $\ref{dis.fig}$ we report  the total population $\avg{n(f)}$
as a function of the  fitness $f=1-\varepsilon$ for a uniform fitness
distribution $\rho(f)=1$ with $f\in(0,1)$.
The figure shows that at low temperature,  in absence of the reproduction
process, high fitness species are not rewarded by the non-neutral
birth-death dynamics. On the other hand, in absence of the negative
selection process, low fitness species are not punished by the non
neutral birth-death dynamics. 
In order to study the non-stationary condensed state, we consider a
fitness distributions of the type $\rho(f)=(\kappa+1)(1-f)^{\kappa}$.

\begin{figure}
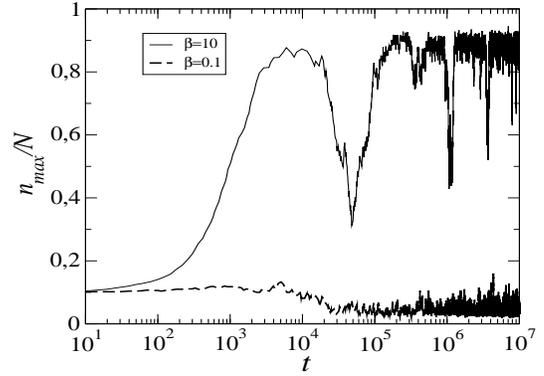

\onefigure[width=70mm,height=50mm]{dyn.eps}
\caption{Time dependence of the fraction of individuals in the
  dominating species below ($\beta=10$) and above ($\beta=0.1$) the
  condensation phase transition. Below the phase transition the
  non-stationary of the process is manifest with the most dominating
  species changing over long time scales. The data  represent a
  single run of the evolutionary dynamics with parameters $N=1000$,
  $\alpha=10$ and $\gamma=0.1$ and $\kappa=0.1$. The critical value of
  the inverse temperature for these parameters is given by $\beta_c=2.4$.}
\label{dyn.fig}
\end{figure}

In fig. $\ref{dyn.fig}$ we report the time dependence of the
fraction of individuals in the most populated species above and below
the condensation transition. Above the condensation phase transition only an infinitesimal fraction of the total population belongs to the most abundant species, while below the condensation transition the dominant species is
populated by a finite fraction of all individuals.
The dominant
species changes over long time scales as it is eventually overcome by other
 fitter species arriving later in the ecosystem.
This process is the process that generalize the principle of
competitive exclusion to open ecologies.
In close ecolgies the competitive exclusion principle in fact states
that at long time scales, the fittest species of the ecosystem is
fixated with higher probability. Here we prove that,  in the presence of
migration,  the ecology might become dominated by a species with high fitness.

\begin{figure}
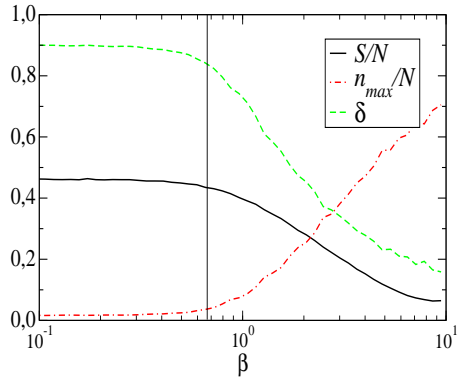

\onefigure[width=60mm,height=50mm]{across1000.eps}
\caption{The normalized number of species $S/N$, the fraction of
  individuals in the dominating species $n_{max}/N$ and the parameter
  $\delta$ which measures the diversity in the population as a function
  of $\beta$ across the
  phase transition for $\alpha=1.5$, $\gamma=1.5$ and $\kappa=1.0$,
  $N=1000$.The solid line indicates the value  of the critical inverse
temperature.}
\label{across}
\end{figure}

In order to characterize the phase transition as a function of
$\beta$ we studied the order parameter $n_{max}/N$ indicating the
fraction of individuals in the dominating species.
Going across the condensation phase transition toward the
non-stationary condensed  state, the diversity of the population rapidly
decreases as it is shown in Fig. $\ref{across}$ where we plot   the
total number of species $S/N$ as a function of $\beta$.
Moreover the inhomogeneity in the systems increases strongly below the
condensation phase transition. In order to measure this inhomogeneity
we plot the parameter $\delta$ defined as
$\delta=-\frac{\ln(Y_2)}{\ln(S)}$ 
with $Y_2$ indicating the participation ratio of the species
population, i.e.
$Y_2=\sum_{i=1}^S \left(\frac{n_i}{N}\right)^2.$

The participation $Y_2$ varies from $Y_2=1/S$ when all the species
are equally populated to $Y_2=1$ when one species dominates the
population. Therefore the parameter $\delta\in(0,1)$ is equal to $\delta=1$ for
equally populated species and is $\delta=0$ when only one species is
occupied. In fig. $\ref{across}$ we show how $n_{max}/N, S/N, $ and $\delta$
varies across the condensation transition as a function of $\beta$.

{\it Conclusions -}
In this paper we have proposed a new, simple, non-neutral stochastic model for 
ecosystems in which species  compete for finite resources. 
In the model  we distinguish different limiting cases:
\begin{itemize}
\item {Case $\beta=0$ -}
We recover the {\it Neutral Model of the Unified Theory of Biodiversity} with species abundance distribution that
can be approximated by a log Fisher distribution with diversity
parameter given by Eq. (\ref{theta}), therefore  $\theta=(\alpha+\gamma)/(1+\alpha+\gamma)$.
\item {Case $\beta>\beta_c$ -}
We have a stationary ecology where the species abundance can be
approximated by a  convolution of log Fisher distributions with
diversity parameter dependent on the fitness $\theta=\theta(f)$ given
by Eq. $(\ref{theta})$ and constants ${\cal Z}_1$ and ${\cal Z}_2$
satisfying Eq.(\ref{selfc}).
\item {Case $\beta<\beta_c$ -}
We have a non stationary ecology with a dominating species and
strongly reduced diversity. 
The  phase transition at $\beta=\beta_c$ extends the  principle of competitive
exclusion  to open ecosystems with high migration rate and shed light on the
instabilities found in presence of high fitness invasive species.
\item{Case $\alpha=0$ and $\beta\rightarrow \infty$ -}
We find a dynamics of extinctions that can be mapped to a mean-field
Bak-Sneppen model and  it can be shown  \cite{prossimo} that this model shows punctuated equilibrium.
\end{itemize}
We believe that the proposed model can provide a general framework
for the study of non-neutral models of biodiversity, presenting in a
unified framework different known limits of stochastic dynamics of ecosystems.
Work on progress is investigating extensions of this model for the description of
ecosystems with non-trivial foodweb prey-predator or mutualistic  
interactions.

\acknowledgments
G.B. acknowledge discussion with Amos Maritan and Kim Sneppen,  support by
the IST STREP GENNETEC contract number 034952 and the  ospitality at
the LPTMS.

\end{document}